\documentclass[aps,a4paper,preprint]{revtex4}%
\usepackage{amsfonts}
\usepackage{amsmath}
\usepackage{amssymb}
\usepackage{graphicx}
\usepackage{setspace}%
\providecommand{\U}[1]{\protect\rule{.1in}{.1in}}
\begin{document}
\title{Self-similar dynamics of air film entrained by a solid disk in confined space:
a simple prototype of topological transitions}
\author{Hana Nakazato, Yuki Yamagishi, and Ko Okumura}
\affiliation{Physics Department, Faculty of Science, Ochanomizu University}
\date{\today}

\begin{abstract}
In hydrodynamic topological transitions, one mass of fluid breaks into two or
two merge into one. For example, in the honey-drop formation when honey
dripping from a spoon, honey is extended to separate into two as the liquid
neck bridging them thins down to micron scales. At the moment when topology
changes due to the breakup, physical observables such as surface curvature
locally diverges. Such singular dynamics have widely attracted physicists,
revealing universality in their self-similar dynamics, which share much in
common with critical phenomena in thermodynamics. Many experimental examples
have been found, which include electric spout and vibration-induced jet
eruption. However, only a few cases have been physically understood on the
basis of equations that govern the singular dynamics and even in such a case
the physical understanding is mathematically complicated inevitably involving
delicate numerical calculations. Here, we study breakup of air film entrained
by a solid disk into viscous liquid in a confined space, which leads to
formation, thinning and breakup of the neck of air. As a result, we
unexpectedly find that equations governing the neck dynamics can be solved
analytically by virtue of two remarkable experimental features: only a single
length scale linearly dependent on time remains near the singularity and
universal scaling functions describing singular neck shape and velocity field
are both analytic. The present solvable case would be essential for our better
understanding of the singular dynamics and will help unveil the physics of
unresolved examples intimately related to daily-life phenomena and diverse
practical applications.

\end{abstract}
\maketitle

The self-similar dynamics in hydrodynamics was already on focus in 1982, when
the dynamics of viscous instability of a moving front was studied
\cite{1992NatureHuppertViscousFlowInstability} and the renormalization group
theory in statistical physics, which elucidates universality appearing in
critical phenomena in thermodynamics, was recognized worldwide beyond fields
\cite{wilson1983renormalization}. In 1993, the self-similar dynamics was first
studied for breakup of droplets (capillary pinch-off)
\cite{1993PREDropBreakupHeleShaw}, one typical example of topological
transitions. In the seminal paper, the scaling ansatz, one of the key ideas
leading to universality in critical phenomena, was shown to be useful. This
paper ignited a surge of publication on the self-similarity in fluid breakup
phenomena in 1993 and 1994
\cite{1993PRLEggersPinchoff,goldstein1993topology,1994ScienceNagelDropFallingFaucet,Eggers1997}%

In 1999, analytical solution for the dynamics of droplet coalescence, another
typical example of hydrodynamic topological transitions, was published
\cite{Eggers1999}. This is followed by many studies on the dynamics of
droplets and bubbles, such as coalescence
\cite{AartsLekkerkerkerGuoWegdamBonn2005,BurtonTaborek2007,paulsen2012inexorable,paulsen2014coalescence,NatCommunNagel2014Coalesce}%
, non-coalescence \cite{RistenpartBirdBelmonteDollarStone2009}, pinch-off
\cite{burton2004PRLpinchoff,burton2008bifurcation}, and droplet impact onto
substrates
\cite{RichardClanetQuere2002,okumura2003water,NagelImpact05,bird2013reducing,liu2015symmetry,Mugele2015NP}%
, with a rapid technological advance in high-speed imaging and numerical
computing. However, since then, the field has rather remotely been developed
from the self-similar dynamics, while one of the main foci has been on simple
scaling laws \cite{CapilaryText}, which is another key concept in
understanding universality in critical phenomena.

Up to date, many types of singular transitions, which is more general than
topological transitions, and associated self-similar dynamics have been found
experimentally in hydrodynamics. Such singular transitions often exhibit
universality via self-similar dynamics near the singular point, while
non-universal behavior is observed under certain conditions
\cite{2003ScienceNagelMemoryDropBreakup}. Other examples often related to
practical applications include spout of liquid jet under electric field
\cite{2000PRLNagelElectricSpout} (closely related to ink-jet printing) and jet
eruption under vibration \cite{2000NatureLathropFluidJetEruption}. In
addition, the self-similar dynamics was found for the first time in droplet
coalescence under electric field \cite{YokotaPNAS2011} (relevant to
microfluidic applications). Self-similarities are sometimes identified not
only in the dynamics as in the above examples but also in the steady-state
flow towards singular transitions, in which topology is not necessarily
changed, as observed in flow-induced air entrainment
\cite{1992JFMMoffattCuspSingularity,2001PRLEggersAirEntrainment,2003PRLEliseFractureLiquid}
and selective withdrawal \cite{2002PRLCohenSelectiveWithdrawal}; both of them
are useful, e.g., for macroscopic \cite{1993BurleyAirEntrain} and microscopic
coatings \cite{2001ScienceCohenSelectiveWithdrawal} important for chemical,
biological, and medical engineering.

As indicated above, each example of singular transitions is often familiar to
everyone and/or important for applications. In fact, our example of fluid
breakup is central to diverse processes, involving two-phase flows,
emulsification and drop formation, in industrial, engineering, and scientific
realms. Understanding physical laws governing the dynamics is useful for
various purposes \cite{frohn2000dynamics}: e.g., (1) industrial production of
food such as mayonnaises and cream and of spray system for painting or
planting, (2) engineering development of microfluidic devices promising for
applications in chemistry, biochemistry and material science, and (3)
atmospheric science dealing with formation of raindrops and thunderstorms.

As for the emergence of universality, singular transitions are analogous to
critical phenomena in thermodynamic transitions
\cite{1972PRLWilsonRG,Cardy,Barenblatt1996scaling}. By identifying
characteristic physical quantities and lengths describing the fluid dynamics
near the singularity, the properties of the interface and the flow can be
understood in terms of a similarity solution near the transition point, which
allows a simplified description of the dynamics, leading to a deeper
understanding of the phenomena.

However, only a few examples of the above-mentioned singular dynamics have
been understood, and even if well understood, the theory is intricate
involving delicate numerical calculations (In contrast, universal dynamics
associated with hydrodynamic instability have been rather simply understood
\cite{1992NatureHuppertViscousFlowInstability,2014ScienceElieGlassyPolymer}).
Because of this, understanding of the singular dynamics is premature. If
otherwise, as has been done for critical phenomena, universalities for the
singular dynamics could have been classified.

Here, we report an experimental and theoretical study of appearance of a
singularity by air entrainment caused by a solid disk into a confined viscous
medium in the Hele-Shaw cell. The remarkable features of the present case are
that only a single length scale remains near the singularity to characterize
the dynamics and that the length scale simply scales linearly with time. These
exceptionally simple characteristics make the theory completely solvable in a
perturbative sense, leading to analytic scaling functions and demonstrating
the essential physical mechanisms of the singular dynamics in a clear way.
Most of previous experimental examples possess more than one remaining length
scales near the singularity and at least one of universal scaling functions is
nonanalytic. This is indeed the case for well-understood cases
\cite{Eggers1997} such as the pioneering study associated with the Hele-Shaw
cell \cite{1993PREDropBreakupHeleShaw} (see just below Eq.(\ref{e3})) and
another seemingly similar case of the flow-induced air entrainment
\cite{1992JFMMoffattCuspSingularity,2003PRLEliseFractureLiquid} (see
Discussion for details). Accordingly, the present study advances our general
understanding of the singular dynamics; it represents an important fundamental
example of the singular dynamics, providing insight into unresolved
self-similar dynamics such as the electric spout of liquid
\cite{2000PRLNagelElectricSpout}, the selective withdrawal
\cite{2002PRLCohenSelectiveWithdrawal}, and the electric-field-induced drop
coalescence \cite{YokotaPNAS2011}, and impacting onto the study of the
dynamics of droplets and bubbles in general.

\section{Results}

\subsection{Experimental}

\begin{figure}[h]
\includegraphics[width=\textwidth]{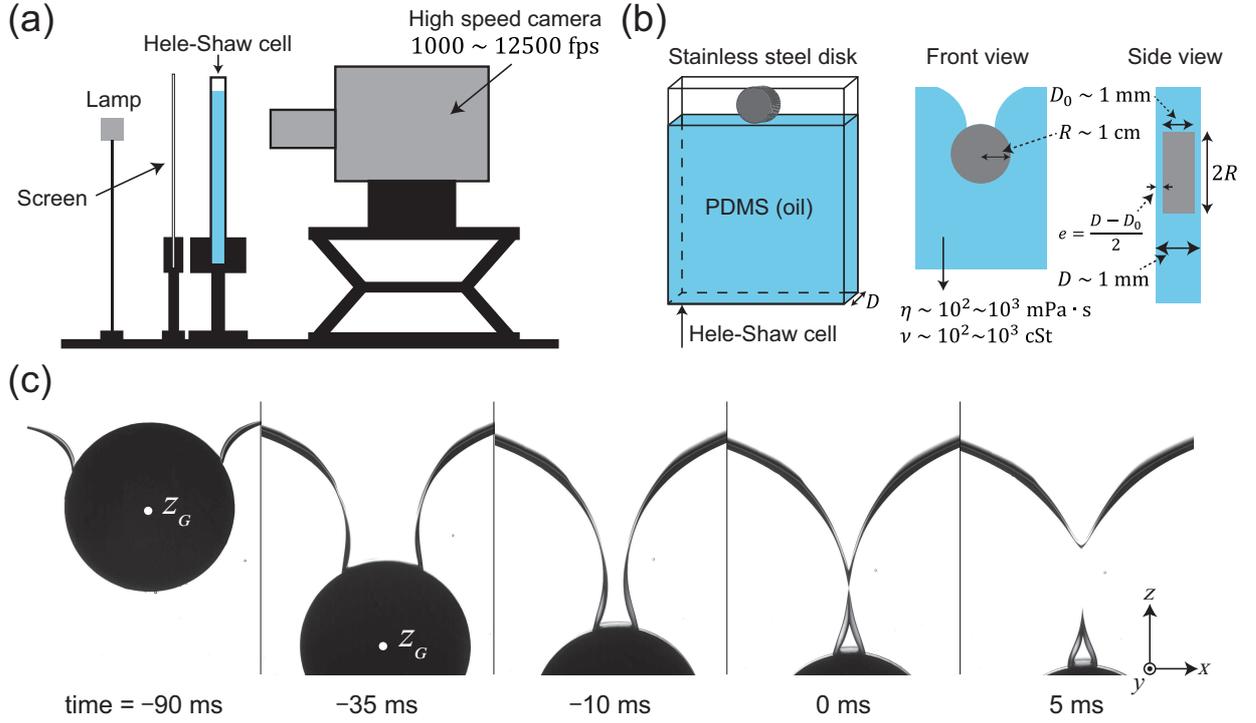}\caption{(a) Experimental setup.
(b) Experimental geometry. The disk of radius $R$ with thickness $D_{0}$ falls
in the Hele-Shaw cell of thickness $D$ filled with a silicone oil of viscosity
$\eta$ and kinematic viscosity $\nu=\eta/\rho$ ($\rho$ is the density of the
oil). Thin oil films of thickness $e$ exist between the surfaces of the disk
and cell walls. (c) Overall pinch-off dynamics of air dragged by the disk
observed from the front of the cell for $(\nu,D,R,e)=(100,4,10,0.5)$ where
$\nu$ is given in the unit cS and the others are in the unit mm.}%
\label{f1}%
\end{figure}

\begin{figure}[h]
\includegraphics[width=\textwidth]{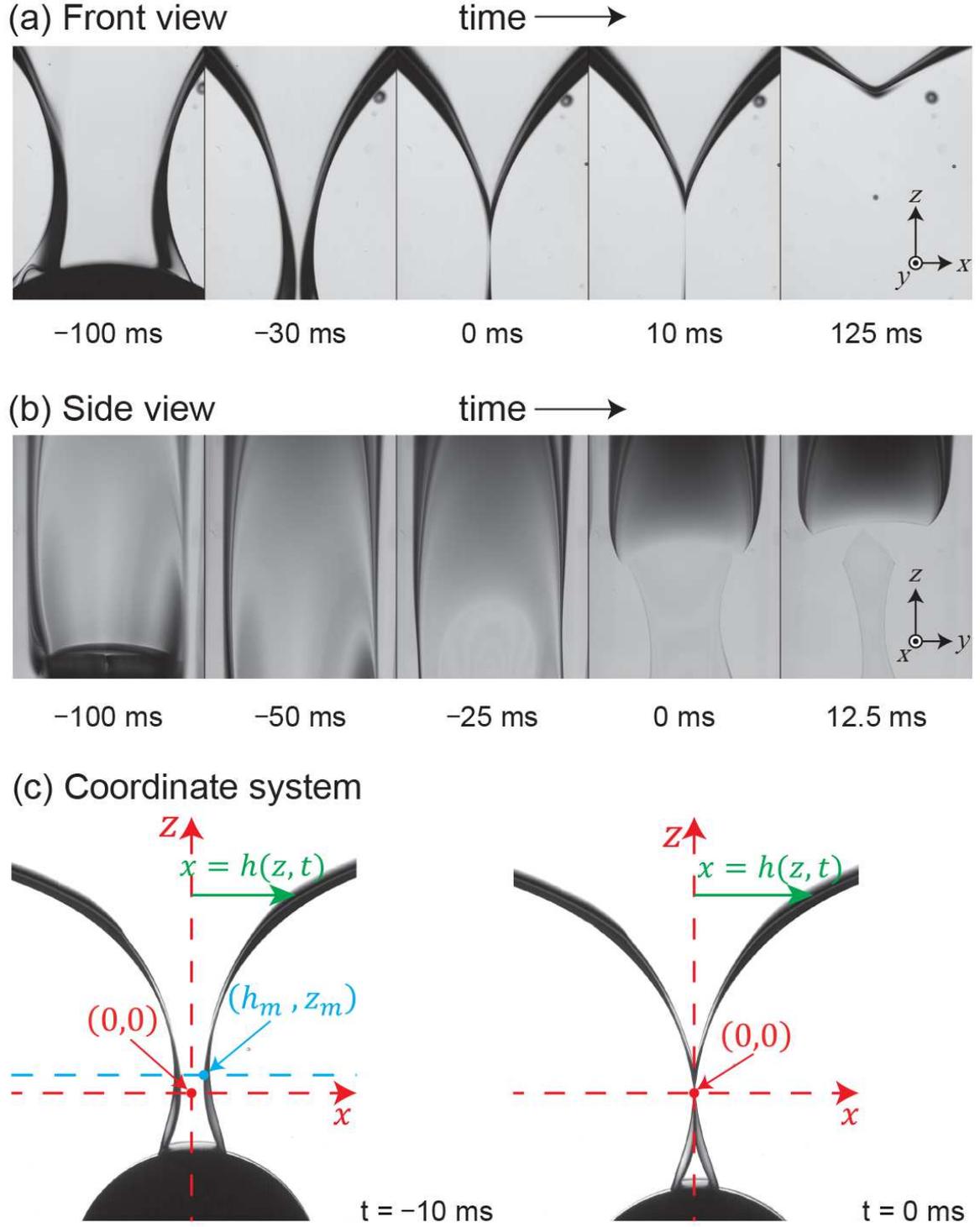}\caption{Pinch-off dynamics
observed from the front (a) and from the side (b) for $(\nu
,D,R,e)=(500,5,10,0.5)$ with the units specified in Fig. \ref{f1}. (c)
Definitions of geometrical parameters, the minimum neck width $2h_{m}$, the
vertical position of the minimum neck $z_{m}$, and the neck profile
$x=h(z,t)$. Snapshots are taken for $(\nu,D,R,e)=(100,3,10,0.5)$.}%
\label{f2}%
\end{figure}

The experimental setup is shown in Fig. \ref{f1}(a). We create a Hele-Shaw
cell with the thickness $D$, and fill the cell with a viscous oil
(polydimethylsiloxane, or PDMS) with the kinematic viscosity $\nu=\eta/\rho$
where $\eta$ and $\rho$ are the viscosity and density of the oil. We insert a
stainless disk [the radius $R$, the thickness $D_{0}$ ($<D$), the density
$\rho_{s}$] at the top of the cell. The bottom of the disk touches the
liquid-air interface with zero velocity and the disk starts to go down in the
oil phase due to gravity. Since the disk thickness is slightly thinner than
the cell thickness $D$, the disk does not directly touch the cell plates and a
thin layer of oil is formed between a front cell plate and a front surface of
the disk. The thickness of the film $e$ is fixed to the value $e=(D-D_{0})/2$
(Fig. \ref{f1}(b)). As a result, the air is dragged by the disk as in Fig.
\ref{f1}(c) forming a singular shape, with details revealed in Fig.
\ref{f2}(a) and (b) (see movies 1 to 4 \cite{SMnakazato}), and finally pinches
off to cause a topological transition.

The ($x,y,z$) coordinate system is specified in Fig. \ref{f2}(c) (see also
Fig. \ref{f1}(c) in which the center of gravity of the disk is denoted by
$z_{G}$). For simplicity, the space-time position at the critical moment of
pinch-off $(t,x,z)=(t_{c},x_{c},z_{c})$ is set to the origin of the $(t,x,y)$
coordinate if not specified. The right-hand side of the air-liquid surface is
described by $x=h(t,z)$. The minimum of the function $h(t,z)$ as a function of
$z$ is denoted as $(x,z)=(h_{m},z_{m})$. Under our experimental conditions,
the shape of the neck is not axisymmetric and is rather independent from the
$y$ coordinate. In fact, as shown in the magnified snapshot in Fig.
\ref{f2}(a) and (b), the singular neck shape near the pinch-point is like a
sheet of thickness $2h_{m}$ and width $\simeq D_{0}$.

Now, we discuss experimental details of the experiment. The height and width
of Hele-Shaw cells are 12 cm and 9 cm, respectively. The two plates are
separated by a spacer of thickness $D$. To make the thickness of two thin film
layers formed on both sides of the disk to be nearly equal, we attach two
small acrylic plates at the inside surfaces of the cell plates near the top of
the cell. The thickness of the small plates are slightly smaller than $e$ and
the attached two plates play a role of a gate for the inserted disk. In the
present experiment, we examined different values of $e$ ($=0.2,0.5,$ and $1.0$
mm) and confirmed that the results are independent of the value of $e$ in the
range. The disk is pre-wetted by the oil before being dropped into the cell;
there is no contact line on the surface of the disk, which removes any effects
of the contact angle on the phenomena. When the cell thickness is smaller than
the values (3 to 5 mm) studied in the present study, the falling velocity of
the disk becomes significantly smaller; there seems to be another regime, in
which the cavity shape near the contact with the disk becomes cone-like,
unlike the present regime characterized by the sheet formation; this issue
will be discussed elsewhere. We used a high-speed camera (Fastcam SA-X,
Photoron) with a macro lens (Micro NIKKOR 60 mm f2.8G ED, Nikon) and analyzed
images with a software (Image J). The density of stainless steel (SUS430) is
7.7 g/cm$^{3}$, while that of PDMS depends on viscosity: 0.965 for 100 cS and
0.97 for 500 and 1000 cS and 0.975 for 10000 cS in the unit g/cm$^{3}$.

\subsection{Dynamics of the neck}

\begin{figure}[h]
\includegraphics[width=\textwidth]{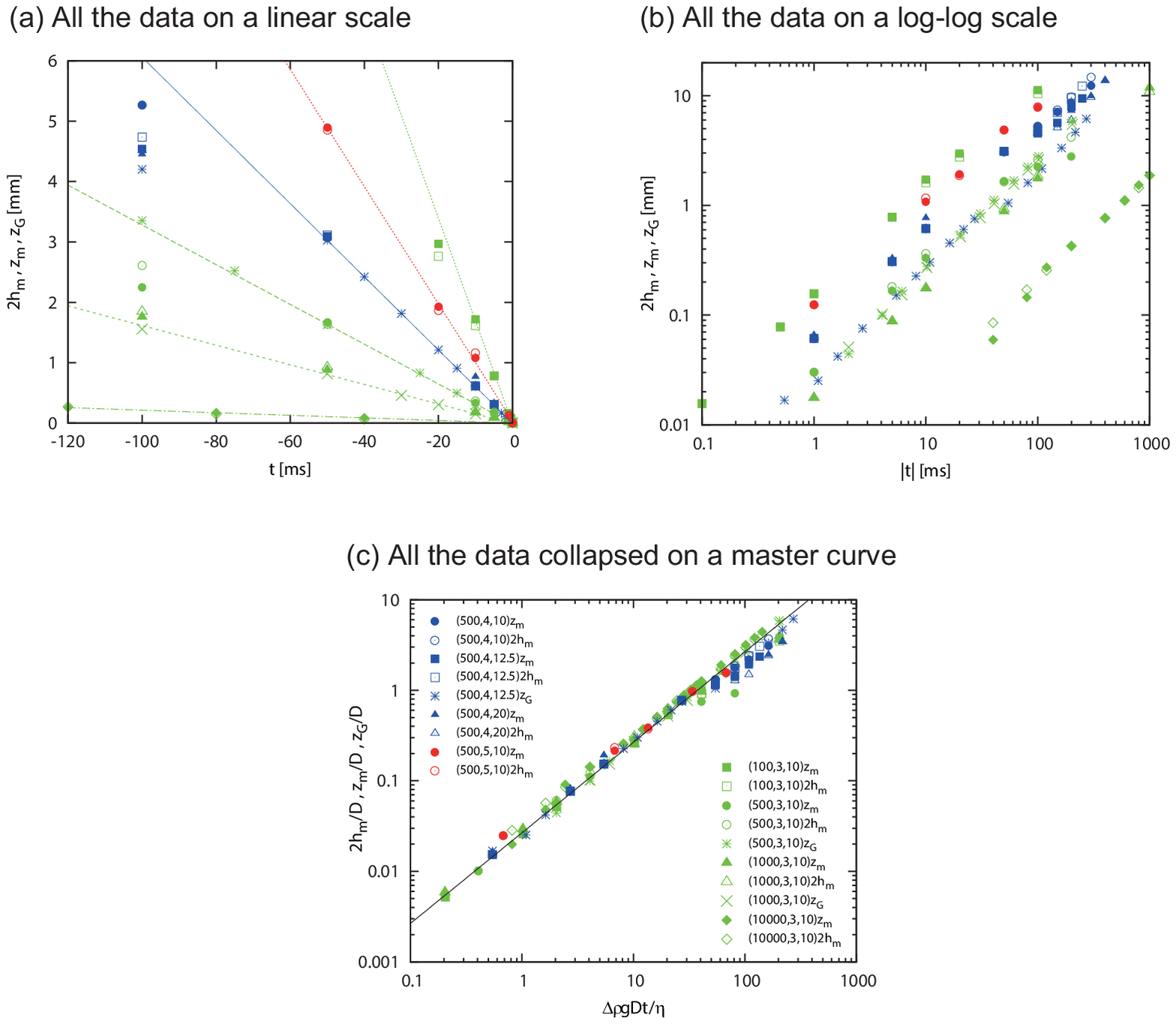}\caption{The neck width $2h_{m}$,
the vertical position of the neck $z_{m}$, and the center of mass of the disk
$z_{G}$ as a function of time $t$. ($t$ and $z_{m}$ are measured from the
origins $t=t_{c}$ and $z=z_{c}$, respectively, while $z_{G}$ from $z=z_{G}$ at
$t=t_{c}$.) The symbols are specified in the plots by the set $(\nu,D,R)$ as
in Figs. \ref{f1} and \ref{f2}. (a) All the data on a linear-linear scale. (b)
The same data on a log-log scale. (c) All the data collapsed on a single
master line with slope one, establishing Eq. (\ref{e1}) with Eq. (\ref{e1b});
the best-fitting line gives $k=0.02764\pm0.00008$. The dashed lines in (a) are
guide for the eye.}%
\label{f3}%
\end{figure}

Figures \ref{f3}(a)-(b) show $z_{m}$ and $h_{m}$, together $z_{G}$, as a
function of time for various parameters $\nu$, $D$, and $R$ ($e=0.5$ mm).
Figure \ref{f3}(c) clearly shows the following relations:%
\begin{equation}
2h_{m}=z_{m}=z_{G}\equiv l(t)\label{e1}%
\end{equation}
where $l(t)$ ($t<0$) is given by%
\begin{equation}
l(t)\equiv v_{0}\times(-t)\text{ \ where }v_{0}=k\frac{\Delta\rho gD^{2}}%
{\eta}\label{e1b}%
\end{equation}
with $k=0.02764\pm0.00008$, which is close to $1/(12\pi)\simeq0.0265$, and
$\Delta\rho=\rho_{s}-\rho$ (see Discussion for the expression for $v_{0}$).

Plots in Fig. \ref{f4}(a)-(c) demonstrate, for two different parameter sets,
that the neck profile $h(t,z)$ has the following scaling form, demonstrating a
self similar dynamics with the single length scale $l(t)$:%
\begin{equation}
2h(t,z)/l(t)=H(z/l(t)) \label{e2}%
\end{equation}
Figure \ref{f4}(d) shows that the data agree quite well with the expression%
\begin{equation}
H(\xi)=1+a(\xi-1)^{2}\text{ with }\xi=z/l \label{e2b}%
\end{equation}
with $a=3.1415\pm0.011$, which is close to $\pi$. As stated above, remaining
of only a single length scale near singularity is an exceptionally simple
case, similar to the coalescence case \cite{YokotaPNAS2011}, for which
physical understanding is lacking.

Further details are discussed below. Since the liquid completely wet the cell
plates, we measure the inner edge of the neck image as $2h$. In Fig. \ref{f3}a
and b, the dashed line (the guide for the eye) are drawn by fitting the first
three data (excluding the data at $t=0$) for $z_{G}$ (or $z_{m}$ if $z_{G}$ is
not available). In Fig. \ref{f4}c and d, the average of the left and right
branches are used for the analysis to reduce the experimental error.

\subsection{Theory}

As shown below, the existence of the self-similar solution in Eq. (\ref{e2})
with its asymptotic form in Eq. (\ref{e2b}) and the linear dependence of
$l(t)$ on $t$ in Eq. (\ref{e1b}) can be explained by considering the air flow
near the neck. Below, the physically most important component of the flow,
i.e., the $z$ direction, $v_{z}$, is denoted as $v$. Considering that a sheet
of air with thickness $\simeq2h$ (and width in the $y$ direction $\simeq
D_{0}$ slightly smaller than $D$) is formed near the neck, as seen in Fig.
\ref{f2}(a) and (b), it is reasonable to assume that $v$ is independent of $y$
near $y=0$ because $h\ll D$. ($y=0$ corresponds to the central position
between the two walls separated by the distance $D$.) We can further assume
that $v$ is independent of $x$, i.e., $v=v_{z}(t,z)$, near the pinch-off
point. This is because the narrowest point of the neck located at
$z=z_{m}=l(t)=-v_{0}t$ moves downwards with the velocity $v_{0}$ ($>0$), and
thus the liquid-air interface drags downwards air at the interface, which
makes a plug flow inside the air neck, since Reynolds number Re $=\rho
_{a}v_{0}h/\eta_{a}$ approaches infinitely small towards the singular point
($\rho_{a}\simeq1$ kg/m$^{3}$ and $\eta_{a}\simeq10^{-5}$ Pa$\cdot$s are the
density and viscosity of air and Re
%TCIMACRO{\TEXTsymbol{<} }%
%BeginExpansion
$<$
%EndExpansion
0.1 for $h<$ 1 mm since $v_{0}<10^{-3}$ m/s). Now that we have shown the
independence of $v$ from $x$ and $y$ near the singularity, we obtain the
following condition
\begin{equation}
v(t,z=z_{m})=-v_{0}, \label{eqa1}%
\end{equation}
and the following equation of continuity:%
\begin{equation}
\frac{\partial h}{\partial t}+\frac{\partial hv}{\partial z}=0. \label{eq1}%
\end{equation}

Eq. (\ref{eq1}) can also be justified in the following manner. The pressure
value at the neck is almost precisely equal to that of the atmospheric
pressure $p_{0}\simeq10^{5}$ Pa, which implies $\partial\rho_{a}/\partial t=0$
(with the air density $\rho_{a}$) even for a thin neck. This is because
Laplace's pressure jump $\Delta p$ is negligible compared with $p_{0}$ as long
as the radius of curvature of the air liquid interface is larger than 0.01 mm,
which is the present case; since the neck forms a sheet, the radius of
curvature is independent of the width of a neck. This is in contrast with the
axisymmetric case, in which the radius of curvature scales as the thin neck
width and thus the pressure jump becomes significant at the neck.

Dynamics of $h$ and $v$ coupled in Eq. (\ref{eq1}) is completely specified by
introducing one more equation: the $z$ component of the Navier-Stokes equation
in the small $h$ limit, in which viscosity dominates inertia (note that
$\partial_{x}v_{z}=0$ as before):%
\begin{equation}
\eta_{a}\frac{\partial^{2}v}{\partial z^{2}}=\frac{\partial\Delta p}{\partial
z}\text{ with }\Delta p=-\gamma\frac{\partial^{2}h}{\partial z^{2}}.
\label{e3}%
\end{equation}
(Here, $\gamma$ is the surface tension of the liquid.) In the pioneering study
\cite{1993PREDropBreakupHeleShaw}, the left-hand side of this equation is
replaced by an averaged Poiseuille flow independent of $z$, which unexpectedly
leads to a much more complicated scenario for the appearance of universality.

Eq. (\ref{e3}) can also be justified in the following manner, i.e., we can
confirm retrospectively that the dominating terms in the $z$ component of the
Navier-Stokes equation is indeed expressed as in Eq. (\ref{e3}) near the
singularity. From the experimental results, near the singular point, we expect
characteristic scales in the $x$ and $z$ directions $l_{x}$ and $l_{z}$ and a
characteristic time scale $\tau$ can be given by $l_{x}=l_{z}=\varepsilon
l_{0}$, and $\tau=\varepsilon t_{0}$ with introducing a small dimensionless
parameter $\varepsilon$ and units of length and time, $l_{0}$ and $t_{0}$,
which is explained as follows. Since the experiment suggests that there is
only a single length scale near the singularity, we can set $l_{x}%
=l_{y}=\varepsilon l_{0}$. The velocity in the $z$ direction is characterized
by $v_{0}$, a finite value independent of $\varepsilon$. This implies the
characteristic time scale $t_{c}$ for the problem is given by $\tau
=\varepsilon t_{0}$ because $v_{0}\simeq l_{x}/\tau\simeq l_{y}/\tau$ is
independent of $\varepsilon$. We can now estimate the order of each term in
the $z$ component of the Navier-Stokes equation. The inertial terms $\rho
_{a}(\partial_{t}v_{z}+v_{x}\partial_{x}v_{z}+v_{z}\partial_{z}v_{z})$ scale
as $\varepsilon^{-1}$ and the gravitational term $\rho_{a}g$ scale as
$\varepsilon^{0}$, and the terms remaining in Eq. (\ref{e3}) all scale as
$\varepsilon^{-2}$, which justifies that Eq. (\ref{e3}) properly collects
relevant leading order terms near the singularity where $\varepsilon\approx0$.

We employ the scaling ansatzs
\cite{1993PREDropBreakupHeleShaw,1993PRLEggersPinchoff,1994ScienceNagelDropFallingFaucet,2003ScienceNagelMemoryDropBreakup,Eggers1997,eggers1995}
\begin{equation}
2h(t,z)=z_{0}\tilde{t}^{\alpha}H(\tilde{z}/\tilde{t}^{\beta});\text{
}v(t,z)=-v_{0}^{\prime}\tilde{t}^{\gamma}V(\tilde{z}/\tilde{t}^{\delta})
\label{an1}%
\end{equation}
with introducing dimensionless scaling function $H$ and $V$ along with the
dimensionless variables \ $\tilde{z}=(z-z_{c})/z_{0}=z/z_{0}$ and \ $\tilde
{t}=v_{0}^{\prime}(t_{c}-t)/z_{0}=(-v_{0}^{\prime}t)/z_{0}$ where $z_{0}$ and
$v_{0}^{\prime}$ are arbitrary length and velocity scales, respectively, and
substitute them into Eqs. (\ref{eq1}) and (\ref{e3}). (Note in Eq. (\ref{an1})
that $\gamma$ denotes a dimensionless exponent, not the surface tension.) We
here require that the ansatzs are relevant for these equations near the
singularity where $\tilde{t}\approx0$ (i.e., equating all the exponents for
the variable $\tilde{t}$) to obtain $\alpha=\beta=\delta=1$ and $\gamma=0$.
This set reproduces (with the identification $v_{0}=v_{0}^{\prime}$) the
scaling form in Eq. (\ref{e2}) and reveals another scaling structure,%

\begin{equation}
v(t,z)=-v_{0}V(z/l(t)) \label{eq2}%
\end{equation}
In other words, the present analysis concludes that there is only a single
length scale $l(t)$ near the singularity as observed in experiment, which
linearly scales with $t$.

Substitution of the scaling forms in Eqs. (\ref{e2}) and (\ref{eq2}), thus
obtained theoretically, into Eq. (\ref{eq1}) makes the variables $t$ and
$\xi=z/l$ to be separated: the original partial differential equation can be
changed into a set of ordinary differential equations (the first equation
confirms the linear $t$-dependence of $l(t)$; see the next paragraph for
details):%
\begin{align}
dl/dt  &  =-v_{0}\label{eq3a}\\
(HV)^{\prime}  &  =\xi H^{\prime}-H \label{eq3b}%
\end{align}
Here, $H$ and $V$ are a function of $\xi$ and the prime indicates the
derivative with respect to $\xi$. The same substitution into Eq. (\ref{e3})
results in the following expression:%
\begin{equation}
V^{\prime\prime}=\lambda H^{\prime\prime\prime}\text{ \ with }\lambda
=-\gamma/(2\eta v_{0}) \label{e3b}%
\end{equation}
Equations (\ref{eq3b}) and (\ref{e3b}) should satisfy the following boundary
conditions (see Fig. \ref{f4} for the first two and Eq. (\ref{eqa1}) for the
last):%
\begin{equation}
H(1)=1\text{, }H^{\prime}(1)=0\text{, }V(1)=1 \label{e3c}%
\end{equation}

Instead of starting from the general ansatzs in Eq. (\ref{an1}), we can start
directly from the scaling ansatz in Eqs. (\ref{e2}) and (\ref{eq2}), motivated
by experimental results, without assuming the linear $t$-dependence of $l(t)$
(with accepting the existence of a single length scale near the singularity as
an experimental fact). From this standpoint, we can show from Eq. (\ref{eq3a})
that $l(t)$ should be linearly dependent on $t$, and we can explain the
asymptotic form of the self similar shape in Eq. (\ref{e2b}).

To our surprise, the boundary problem defined by the two ordinary differential
equations for $H$ and $V$ in Eqs. (\ref{eq3b}) and (\ref{e3b}) with the
boundary conditions in Eq. (\ref{e3c}) can be solved perturbatively as
explained in the next paragraph. By assuming the series expansion
$H(\xi)=a_{0}+a_{1}\zeta+a_{2}\zeta^{2}+\cdots$ with $\zeta=\xi-1$, we can
show $a_{0}=1$, $a_{1}=0$, $a_{2}=3\lambda a_{4}$, $a_{3}=0,\cdots$, i.e., for
$\zeta\ll1$
\begin{equation}
H(\xi)=1+a_{2}\zeta^{2}+\cdots,\text{ and }V(\xi)=1-\zeta+\cdots, \label{e4}%
\end{equation}
which explains the experimental result including Eq. (\ref{e2b}). Note that we
can show $a_{n}=0$ for odd $n$ and we can derive expression for $a_{n}$ as a
function of $a_{m}$ (with $m<n$) for even $n$ up to any desired order: the
present boundary problem can be solved completely in a perturbative sense. As
far as we know, there have been no previous examples of hydrodynamic singular
transition in which scaling functions for the shape and velocity ($H$ and $V$)
are both analytic; in all the previously known examples, at least one of the
scaling functions is nonanalytic.

Now we derive Eq. (\ref{e4}). Equation (\ref{eq3b}) can be rearranged as%
\begin{equation}
(1+V^{\prime})H=(\xi-V)H^{\prime},
\end{equation}
from which we obtain%
\begin{equation}
\frac{dH}{H}=f(\xi)d\xi\text{ with }f(\xi)=\frac{1+V^{\prime}}{\xi-V}
\label{eqq1b}%
\end{equation}
By using the boundary condition $H(1)=1$, we get%
\begin{equation}
H(\xi)=e^{F(\xi)}\text{ with }F(\xi)=\int_{1}^{\xi}f(\xi)d\xi\text{ }
\label{eqq1}%
\end{equation}
Since the definition of $f(\xi)$ in Eq. (\ref{eqq1b}) can be recast into the
form%
\begin{equation}
V^{\prime}+f(\xi)V=\xi f(\xi)-1,
\end{equation}
This first-order linear differential equation for $V$ can be solved as%
\begin{equation}
V(\xi)=e^{-F(\xi)}\left[  \int_{1}^{\xi}d\tilde{\xi}e^{F(\tilde{\xi})}\left\{
\tilde{\xi}f(\tilde{\xi})-1\right\}  +1\right]  ,
\end{equation}
which satisfies $V(1)=1$. With using Eq. (\ref{eqq1}), we get%
\begin{align}
V(\xi)  &  =\frac{1}{H}\left[  \int_{1}^{\xi}d\tilde{\xi}H\left\{  \tilde{\xi
}\frac{H^{\prime}}{H}-1\right\}  +1\right] \\
&  =\xi-2\int_{1}^{\xi}d\tilde{\xi}H(\tilde{\xi})/H(\xi)
\end{align}
From the conditions $H(1)=1$ and $H^{\prime}(1)=0$, we expand $H(\xi)$ as%
\begin{equation}
H(\xi)=1+a_{2}\zeta^{2}+a_{3}\zeta^{3}+a_{4}\zeta^{4}+\cdots\label{eqq2}%
\end{equation}
with $\zeta=\xi-1$. From this we obtain%
\begin{align}
V(\xi)  &  =\xi-2\zeta\frac{1+a_{2}\zeta^{2}/3+a_{3}\zeta^{3}/4+a_{4}\zeta
^{4}/5+\cdots}{1+a_{2}\zeta^{2}+a_{3}\zeta^{3}+a_{4}\zeta^{4}+\cdots
}\label{eqq3}\\
&  =1-\zeta+2\zeta(2a_{2}\zeta^{2}/3+3a_{3}\zeta^{3}/4-(2a_{2}^{2}%
/3-4a_{4}/5)\zeta^{4}+\cdots)
\end{align}
Substituting Eqs. (\ref{eqq2}) and (\ref{eqq3}) into Eq. (\ref{e3b}), we can
determine $a_{n}~$perturbatively as a function of $a_{m}$ (with $m<n$) to
obtain Eq. (\ref{e4}).

\section{Discussion}

Among the previous studies of singular transitions, the present problem may be
very similar to air entrainment induced by flow
\cite{1992JFMMoffattCuspSingularity,2003PRLEliseFractureLiquid} in that (1)
this singularity occurs for air surrounded by viscous liquid and (2) a thin
sheet of air is formed so that the problem is two-dimensional but lacks axial
symmetry. However, there is an essential difference: the present
self-similarity is for the dynamics, while the self-similarity discussed for
the flow-induced air entrainment is for the steady-state flow approaching a
transition. In fact, the universal scaling function for the shape in the
previous study cannot be expressed as a Taylor series expansion but needs a
Frobenius series with a singular exponent $3/2$.

We can further explain why $v_{0\text{ }}$is given by the second expression in
Eq. (\ref{e1b}). For this purpose, we note that $v_{0}$ is the magnitude of
falling velocity $U$ $(=dz_{G}/dt)$ of the disk in the oil phase, as seen from
Eqs. (\ref{e1}) and (\ref{e1b}). The falling velocity $U$ could be determined
by the balance between the gravitational energy gain $\Delta\rho gR^{2}D_{0}U$
per time and an appropriate viscous dissipation, which should be the most
dominant one among the following three \cite{okumura2017AdvCI}: dissipation
associated with the Couette flow developed inside thin films between the disk
surface and the cell wall $\simeq\eta(U/e)^{2}R^{2}e$ and dissipations
associated with Poiseuille flows around the disk corresponding to the velocity
gradient in the $y$ direction $\simeq\eta(U/D)^{2}R^{2}D$ and in the radial
direction for the disk $\simeq\eta(U/R)^{2}R^{2}D$. The last dissipation may
be smaller than the first two because $e,D\ll R$. However, the relative
importance of the first and second is delicate; while the volume of
dissipation for the first is well described by $2\pi R^{2}D$, that for the
second can be $\pi(cR)^{2}D$ with a fairly large numerical constant $c$. In
the present case, $c$ seems indeed fairly large and the second dissipation
seems the most dominant. This is because the balance of this dissipation and
the gravitational energy gain gives $U\simeq\Delta\rho gD^{2}/\eta$, which is
consistent with Eq. (\ref{e1b}).

\begin{figure}[h]
\includegraphics[width=\textwidth]{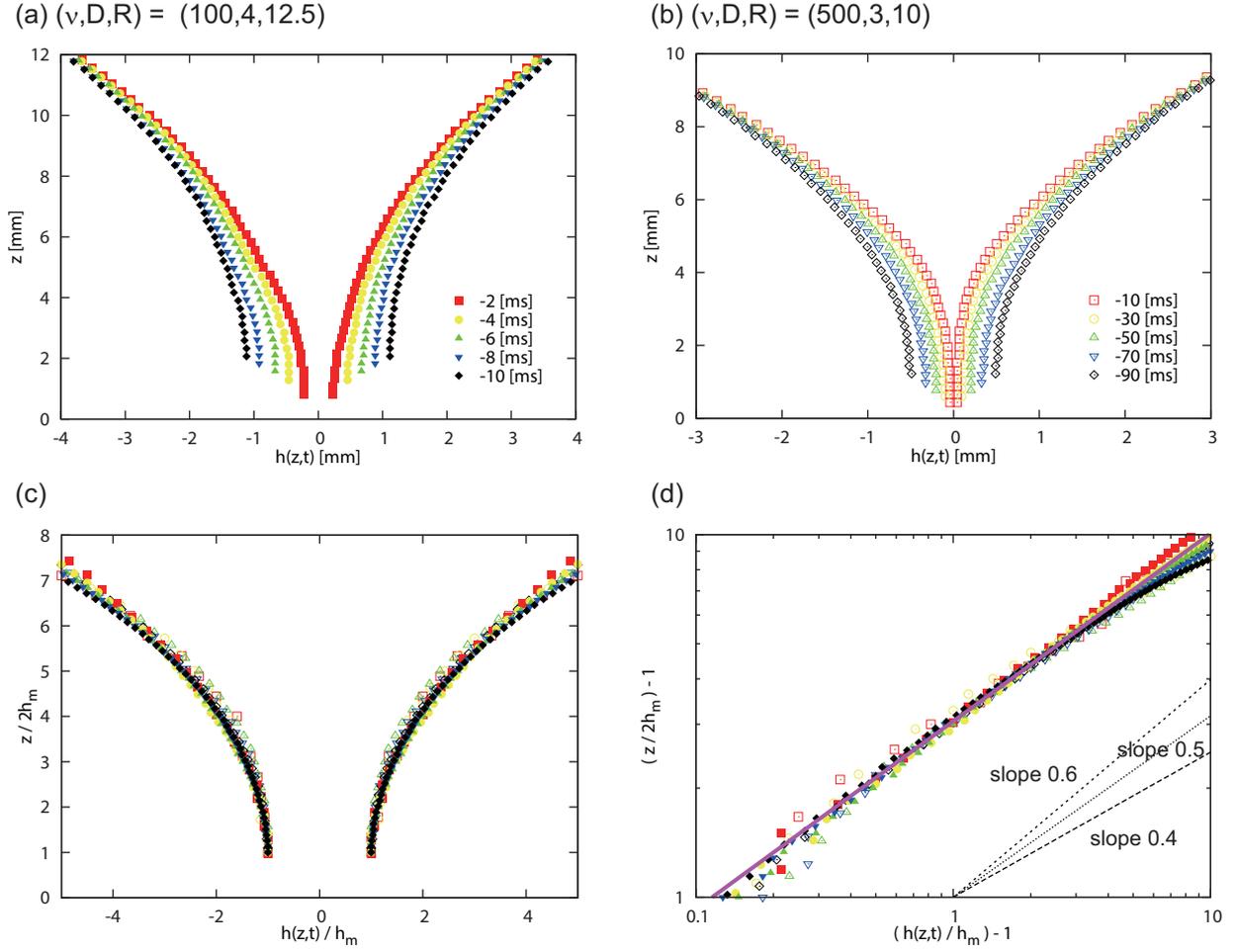}\caption{Self-similar dynamics of
the neck shape. (a) and (b) Neck profile $h(z,t)$ as a function of $z~$at
different times for $(\nu,D,R,e)=(100,4,12.5,0.5)$ and $(500,3,10,0.5)$ in the
units cS and mm. (c) Profiles in (a) and (b) collapsed by rescaling with a
single length scale $2h_{m}=z_{m}=l(t)$. (d) Right branches of the collapsed
neck shape in (c) on a log-log scale, establishing Eq. (\ref{e2b}) with the
best-fitting line giving $a=3.1415\pm0.011$.}%
\label{f4}%
\end{figure}

In summary, our theoretical result explains experimental findings, starting
from the governing equations in (\ref{eq1}) and (\ref{e3}); by virtue of the
ansatzs in Eq. (\ref{an1}), we showed that only a single length scale remains
near the singularity and this length linearly scales with $t$, and we further
explained why we observed the self similar shape dynamics demonstrated in Fig.
\ref{f4} and represented by Eqs. (\ref{e2}) and (\ref{e2b}).

\begin{acknowledgments}
This work was partly supported by Grant-in-Aid for Scientific Research (A)
(No. 24244066) of JSPS, Japan, and by ImPACT Program of Council for Science,
Technology and Innovation (Cabinet Office, Government of Japan).
\end{acknowledgments}

%\bibliographystyle{naturemagMy}
%\bibliography{C:/Users/okumura/Documents/main/JabRef/granular,C:/Users/okumura/Documents/main/JabRef/fracture,C:/Users/okumura/Documents/main/JabRef/wetting}

\begin{thebibliography}{10}
\expandafter\ifx\csname url\endcsname\relax
  \def\url#1{\texttt{#1}}\fi
\expandafter\ifx\csname urlprefix\endcsname\relax\def\urlprefix{URL }\fi
\providecommand{\bibinfo}[2]{#2}
\providecommand{\eprint}[2][]{\url{#2}}

\bibitem{1992NatureHuppertViscousFlowInstability}
\bibinfo{author}{Huppert, H.~E.}
\newblock \bibinfo{title}{Flow and instability of a viscous current down a
  slope}.
\newblock \emph{\bibinfo{journal}{Nature}} \textbf{\bibinfo{volume}{300}},
  \bibinfo{pages}{427--429} (\bibinfo{year}{1982}).

\bibitem{wilson1983renormalization}
\bibinfo{author}{Wilson, K.~G.}
\newblock \bibinfo{title}{The renormalization group and critical phenomena}.
\newblock \emph{\bibinfo{journal}{Reviews of Modern Physics}}
  \textbf{\bibinfo{volume}{55}}, \bibinfo{pages}{583} (\bibinfo{year}{1983}).

\bibitem{1993PREDropBreakupHeleShaw}
\bibinfo{author}{Constantin, P.} \emph{et~al.}
\newblock \bibinfo{title}{Droplet breakup in a model of the hele-shaw cell}.
\newblock \emph{\bibinfo{journal}{Phys. Rev. E}} \textbf{\bibinfo{volume}{47}},
  \bibinfo{pages}{4169--4181} (\bibinfo{year}{1993}).

\bibitem{1993PRLEggersPinchoff}
\bibinfo{author}{Eggers, J.}
\newblock \bibinfo{title}{Universal pinching of 3d axisymmetric free-surface
  flow}.
\newblock \emph{\bibinfo{journal}{Phys. Rev. Lett.}}
  \textbf{\bibinfo{volume}{71}}, \bibinfo{pages}{3458} (\bibinfo{year}{1993}).

\bibitem{goldstein1993topology}
\bibinfo{author}{Goldstein, R.~E.}, \bibinfo{author}{Pesci, A.~I.} \&
  \bibinfo{author}{Shelley, M.~J.}
\newblock \bibinfo{title}{Topology transitions and singularities in viscous
  flows}.
\newblock \emph{\bibinfo{journal}{Physical Review Letters}}
  \textbf{\bibinfo{volume}{70}}, \bibinfo{pages}{3043} (\bibinfo{year}{1993}).

\bibitem{1994ScienceNagelDropFallingFaucet}
\bibinfo{author}{Shi, X.}, \bibinfo{author}{Brenner, M.} \&
  \bibinfo{author}{Nagel, S.}
\newblock \bibinfo{title}{A cascade of structure in a drop falling from a
  faucet}.
\newblock \emph{\bibinfo{journal}{Science}} \textbf{\bibinfo{volume}{265}},
  \bibinfo{pages}{219} (\bibinfo{year}{1994}).

\bibitem{Eggers1997}
\bibinfo{author}{Eggers, J.}
\newblock \bibinfo{title}{Nonlinear dynamics and breakup of free-surface
  flows}.
\newblock \emph{\bibinfo{journal}{Rev. Mod. Phys.}}
  \textbf{\bibinfo{volume}{69}}, \bibinfo{pages}{865--930}
  (\bibinfo{year}{1997}).

\bibitem{Eggers1999}
\bibinfo{author}{Eggers, J.}, \bibinfo{author}{Lister, J.} \&
  \bibinfo{author}{Stone, H.}
\newblock \bibinfo{title}{Coalescence of liquid drops}.
\newblock \emph{\bibinfo{journal}{J. Fluid Mech.}}
  \textbf{\bibinfo{volume}{401}}, \bibinfo{pages}{293--310}
  (\bibinfo{year}{1999}).

\bibitem{AartsLekkerkerkerGuoWegdamBonn2005}
\bibinfo{author}{Aarts, D. G. A.~L.}, \bibinfo{author}{Lekkerkerker, H. N.~W.},
  \bibinfo{author}{Guo, H.}, \bibinfo{author}{Wegdam, G.~H.} \&
  \bibinfo{author}{Bonn, D.}
\newblock \bibinfo{title}{Hydrodynamics of droplet coalescence}.
\newblock \emph{\bibinfo{journal}{Phys. Rev. Lett.}}
  \textbf{\bibinfo{volume}{95}}, \bibinfo{pages}{164503}
  (\bibinfo{year}{2005}).

\bibitem{BurtonTaborek2007}
\bibinfo{author}{Burton, J.~C.} \& \bibinfo{author}{Taborek, P.}
\newblock \bibinfo{title}{Role of dimensionality and axisymmetry in fluid
  pinch-off and coalescence}.
\newblock \emph{\bibinfo{journal}{Phys. Rev. Lett.}}
  \textbf{\bibinfo{volume}{98}}, \bibinfo{pages}{224502}
  (\bibinfo{year}{2007}).

\bibitem{paulsen2012inexorable}
\bibinfo{author}{Paulsen, J.~D.} \emph{et~al.}
\newblock \bibinfo{title}{The inexorable resistance of inertia determines the
  initial regime of drop coalescence}.
\newblock \emph{\bibinfo{journal}{Proceedings of the National Academy of
  Sciences}} \textbf{\bibinfo{volume}{109}}, \bibinfo{pages}{6857--6861}
  (\bibinfo{year}{2012}).

\bibitem{paulsen2014coalescence}
\bibinfo{author}{Paulsen, J.~D.}, \bibinfo{author}{Carmigniani, R.},
  \bibinfo{author}{Kannan, A.}, \bibinfo{author}{Burton, J.~C.} \&
  \bibinfo{author}{Nagel, S.~R.}
\newblock \bibinfo{title}{Coalescence of bubbles and drops in an outer fluid}.
\newblock \emph{\bibinfo{journal}{Nature communications}}
  \textbf{\bibinfo{volume}{5}}, \bibinfo{pages}{3182} (\bibinfo{year}{2014}).

\bibitem{NatCommunNagel2014Coalesce}
\bibinfo{author}{Paulsen, J.~D.}, \bibinfo{author}{Carmigniani, R.},
  \bibinfo{author}{Kannan, A.}, \bibinfo{author}{Burton, J.~C.} \&
  \bibinfo{author}{Nagel, S.~R.}
\newblock \bibinfo{title}{Coalescence of bubbles and drops in an outer fluid}.
\newblock \emph{\bibinfo{journal}{Nature Commun.}} \textbf{\bibinfo{volume}{5}}
  (\bibinfo{year}{2014}).

\bibitem{RistenpartBirdBelmonteDollarStone2009}
\bibinfo{author}{Ristenpart, W.~D.}, \bibinfo{author}{Bird, J.~C.},
  \bibinfo{author}{Belmonte, A.}, \bibinfo{author}{Dollar, F.} \&
  \bibinfo{author}{Stone, H.~A.}
\newblock \bibinfo{title}{Non-coalescence of oppositely charged drops}.
\newblock \emph{\bibinfo{journal}{Nature}} \textbf{\bibinfo{volume}{461}},
  \bibinfo{pages}{377--380} (\bibinfo{year}{2009}).

\bibitem{burton2004PRLpinchoff}
\bibinfo{author}{Burton, J.}, \bibinfo{author}{Rutledge, J.} \&
  \bibinfo{author}{Taborek, P.}
\newblock \bibinfo{title}{Fluid pinch-off dynamics at nanometer length scales}.
\newblock \emph{\bibinfo{journal}{Physical review letters}}
  \textbf{\bibinfo{volume}{92}}, \bibinfo{pages}{244505}
  (\bibinfo{year}{2004}).

\bibitem{burton2008bifurcation}
\bibinfo{author}{Burton, J.} \& \bibinfo{author}{Taborek, P.}
\newblock \bibinfo{title}{Bifurcation from bubble to droplet behavior in
  inviscid pinch-off}.
\newblock \emph{\bibinfo{journal}{Physical review letters}}
  \textbf{\bibinfo{volume}{101}}, \bibinfo{pages}{214502}
  (\bibinfo{year}{2008}).

\bibitem{RichardClanetQuere2002}
\bibinfo{author}{Richard, D.}, \bibinfo{author}{Clanet, C.} \&
  \bibinfo{author}{Qu\'{e}r\'{e}, D.}
\newblock \bibinfo{title}{Surface phenomena: Contact time of a bouncing drop}.
\newblock \emph{\bibinfo{journal}{Nature}} \textbf{\bibinfo{volume}{417}},
  \bibinfo{pages}{811--} (\bibinfo{year}{2002}).

\bibitem{okumura2003water}
\bibinfo{author}{Okumura, K.}, \bibinfo{author}{Chevy, F.},
  \bibinfo{author}{Richard, D.}, \bibinfo{author}{Qu{\'e}r{\'e}, D.} \&
  \bibinfo{author}{Clanet, C.}
\newblock \bibinfo{title}{Water spring: A model for bouncing drops}.
\newblock \emph{\bibinfo{journal}{EPL (Europhysics Letters)}}
  \textbf{\bibinfo{volume}{62}}, \bibinfo{pages}{237} (\bibinfo{year}{2003}).

\bibitem{NagelImpact05}
\bibinfo{author}{Xu, L.}, \bibinfo{author}{Zhang, W.~W.} \&
  \bibinfo{author}{Nagel, S.~R.}
\newblock \bibinfo{title}{Drop splashing on a dry smooth surface}.
\newblock \emph{\bibinfo{journal}{Phys. Rev. Lett.}}
  \textbf{\bibinfo{volume}{94}}, \bibinfo{pages}{184505}
  (\bibinfo{year}{2005}).

\bibitem{bird2013reducing}
\bibinfo{author}{Bird, J.~C.}, \bibinfo{author}{Dhiman, R.},
  \bibinfo{author}{Kwon, H.-M.} \& \bibinfo{author}{Varanasi, K.~K.}
\newblock \bibinfo{title}{Reducing the contact time of a bouncing drop}.
\newblock \emph{\bibinfo{journal}{Nature}} \textbf{\bibinfo{volume}{503}},
  \bibinfo{pages}{385--388} (\bibinfo{year}{2013}).

\bibitem{liu2015symmetry}
\bibinfo{author}{Liu, Y.}, \bibinfo{author}{Andrew, M.}, \bibinfo{author}{Li,
  J.}, \bibinfo{author}{Yeomans, J.~M.} \& \bibinfo{author}{Wang, Z.}
\newblock \bibinfo{title}{Symmetry breaking in drop bouncing on curved
  surfaces}.
\newblock \emph{\bibinfo{journal}{Nature communications}}
  \textbf{\bibinfo{volume}{6}} (\bibinfo{year}{2015}).

\bibitem{Mugele2015NP}
\bibinfo{author}{De~Ruiter, J.}, \bibinfo{author}{Lagraauw, R.},
  \bibinfo{author}{Van Den~Ende, D.} \& \bibinfo{author}{Mugele, F.}
\newblock \bibinfo{title}{Wettability-independent bouncing on flat surfaces
  mediated by thin air films}.
\newblock \emph{\bibinfo{journal}{Nature physics}}
  \textbf{\bibinfo{volume}{11}}, \bibinfo{pages}{48--53}
  (\bibinfo{year}{2015}).

\bibitem{CapilaryText}
\bibinfo{author}{de~Gennes, P.-G.}, \bibinfo{author}{Brochard-Wyart, F.} \&
  \bibinfo{author}{Qu\'{e}r\'{e}, D.}
\newblock \emph{\bibinfo{title}{Gouttes, Bulles, Perles et Ondes, 2nd. eds.}}
  (\bibinfo{publisher}{Belin, Paris}, \bibinfo{year}{2005}).

\bibitem{2003ScienceNagelMemoryDropBreakup}
\bibinfo{author}{Doshi, P.} \emph{et~al.}
\newblock \bibinfo{title}{Persistence of memory in drop breakup: The breakdown
  of universality}.
\newblock \emph{\bibinfo{journal}{Science}} \textbf{\bibinfo{volume}{302}},
  \bibinfo{pages}{1185--1188} (\bibinfo{year}{2003}).

\bibitem{2000PRLNagelElectricSpout}
\bibinfo{author}{Oddershede, L.} \& \bibinfo{author}{Nagel, S.~R.}
\newblock \bibinfo{title}{Singularity during the onset of an
  electrohydrodynamic spout}.
\newblock \emph{\bibinfo{journal}{Phys. Rev. Lett.}}
  \textbf{\bibinfo{volume}{85}}, \bibinfo{pages}{1234--1237}
  (\bibinfo{year}{2000}).

\bibitem{2000NatureLathropFluidJetEruption}
\bibinfo{author}{Zeff, B.~W.}, \bibinfo{author}{Kleber, B.},
  \bibinfo{author}{Fineberg, J.} \& \bibinfo{author}{Lathrop, D.~P.}
\newblock \bibinfo{title}{Singularity dynamics in curvature collapse and jet
  eruption on a fluid surface}.
\newblock \emph{\bibinfo{journal}{Nature}} \textbf{\bibinfo{volume}{403}},
  \bibinfo{pages}{401--404} (\bibinfo{year}{2000}).

\bibitem{YokotaPNAS2011}
\bibinfo{author}{Yokota, M.} \& \bibinfo{author}{Okumura, K.}
\newblock \bibinfo{title}{Dimensional crossover in the coalescence dynamics of
  viscous drops confined in between two plates}.
\newblock \emph{\bibinfo{journal}{Proc. Nat. Acad. Sci. (U.S.A.)}}
  \textbf{\bibinfo{volume}{108}}, \bibinfo{pages}{6395--6398; In this issue,
  PNAS, 108 (2011) 6337.} (\bibinfo{year}{2011}).

\bibitem{1992JFMMoffattCuspSingularity}
\bibinfo{author}{Jeong, J.-T.} \& \bibinfo{author}{Moffatt, H.}
\newblock \bibinfo{title}{Free-surface cusps associated with flow at low
  reynolds number}.
\newblock \emph{\bibinfo{journal}{J. Fluid Mech.}}
  \textbf{\bibinfo{volume}{241}}, \bibinfo{pages}{1--22}
  (\bibinfo{year}{1992}).

\bibitem{2001PRLEggersAirEntrainment}
\bibinfo{author}{Eggers, J.}
\newblock \bibinfo{title}{Air entrainment through free-surface cusps}.
\newblock \emph{\bibinfo{journal}{Phys. Rev. Lett.}}
  \textbf{\bibinfo{volume}{86}}, \bibinfo{pages}{4290} (\bibinfo{year}{2001}).

\bibitem{2003PRLEliseFractureLiquid}
\bibinfo{author}{Lorenceau, {\'E}.}, \bibinfo{author}{Restagno, F.} \&
  \bibinfo{author}{Qu{\'e}r{\'e}, D.}
\newblock \bibinfo{title}{Fracture of a viscous liquid}.
\newblock \emph{\bibinfo{journal}{Phys. Rev. Lett.}}
  \textbf{\bibinfo{volume}{90}}, \bibinfo{pages}{184501}
  (\bibinfo{year}{2003}).

\bibitem{2002PRLCohenSelectiveWithdrawal}
\bibinfo{author}{Cohen, I.} \& \bibinfo{author}{Nagel, S.~R.}
\newblock \bibinfo{title}{Scaling at the selective withdrawal transition
  through a tube suspended above the fluid surface}.
\newblock \emph{\bibinfo{journal}{Phys. Rev. Lett.}}
  \textbf{\bibinfo{volume}{88}}, \bibinfo{pages}{074501}
  (\bibinfo{year}{2002}).

\bibitem{1993BurleyAirEntrain}
\bibinfo{author}{Burley, R.}
\newblock \bibinfo{title}{Mechanism and mechanics of air entrainment in coating
  processes}.
\newblock \emph{\bibinfo{journal}{Spec. Publ. R. Soc. Chem.}}
  \textbf{\bibinfo{volume}{129}}, \bibinfo{pages}{94--107}
  (\bibinfo{year}{1993}).

\bibitem{2001ScienceCohenSelectiveWithdrawal}
\bibinfo{author}{Cohen, I.}, \bibinfo{author}{Li, H.},
  \bibinfo{author}{Hougland, J.~L.}, \bibinfo{author}{Mrksich, M.} \&
  \bibinfo{author}{Nagel, S.~R.}
\newblock \bibinfo{title}{Using selective withdrawal to coat microparticles}.
\newblock \emph{\bibinfo{journal}{Science}} \textbf{\bibinfo{volume}{292}},
  \bibinfo{pages}{265--267} (\bibinfo{year}{2001}).

\bibitem{frohn2000dynamics}
\bibinfo{author}{Frohn, A.} \& \bibinfo{author}{Roth, N.}
\newblock \emph{\bibinfo{title}{Dynamics of droplets}}
  (\bibinfo{publisher}{Springer Science \& Business Media},
  \bibinfo{year}{2000}).

\bibitem{1972PRLWilsonRG}
\bibinfo{author}{Wilson, K.~G.}
\newblock \bibinfo{title}{Feynman-graph expansion for critical exponents}.
\newblock \emph{\bibinfo{journal}{Phys. Rev. Lett.}}
  \textbf{\bibinfo{volume}{28}}, \bibinfo{pages}{548} (\bibinfo{year}{1972}).

\bibitem{Cardy}
\bibinfo{author}{Cardy, J.}
\newblock \emph{\bibinfo{title}{Scaling and Renormalization in Statistical
  Physics}} (\bibinfo{publisher}{Cambridge Univ. Press, Cambridge},
  \bibinfo{year}{1996}).

\bibitem{Barenblatt1996scaling}
\bibinfo{author}{Barenblatt, G.~I.}
\newblock \emph{\bibinfo{title}{Scaling, self-similarity, and intermediate
  asymptotics: dimensional analysis and intermediate asymptotics}},
  vol.~\bibinfo{volume}{14} (\bibinfo{publisher}{Cambridge University Press},
  \bibinfo{year}{1996}).

\bibitem{2014ScienceElieGlassyPolymer}
\bibinfo{author}{Chai, Y.} \emph{et~al.}
\newblock \bibinfo{title}{A direct quantitative measure of surface mobility in
  a glassy polymer}.
\newblock \emph{\bibinfo{journal}{Science}} \textbf{\bibinfo{volume}{343}},
  \bibinfo{pages}{994--999} (\bibinfo{year}{2014}).

\bibitem{SMnakazato}
\bibinfo{title}{See supplemental material at [url will be inserted by
  publisher] for movies 1 to 4.} .

\bibitem{eggers1995}
\bibinfo{author}{Eggers, J.}
\newblock \bibinfo{title}{Theory of drop formation}.
\newblock \emph{\bibinfo{journal}{Physics of Fluids}}
  \textbf{\bibinfo{volume}{7}}, \bibinfo{pages}{941--953}
  (\bibinfo{year}{1995}).

\bibitem{okumura2017AdvCI}
\bibinfo{author}{Okumura, K.}
\newblock \bibinfo{title}{Viscous dynamics of drops and bubbles in hele-shaw
  cells: drainage, drag friction, coalescence, and bursting}.
\newblock \emph{\bibinfo{journal}{Advances in Colloid and Interface Science;
  https://doi.org/10.1016/j.cis.2017.07.021}}  (\bibinfo{year}{2017}).

\end{thebibliography}

\clearpage

\begin{widetext}
\section*{Figures}
%Notice that the order of figures are changed.
\end{widetext}

\pagebreak

\end{document}